\documentclass[preprint2,numberedappendix,iop]{emulateapj-rtx4}
\usepackage{graphicx}
\usepackage{bm}
\usepackage{url}
\usepackage{times}
\usepackage{amsmath}
\usepackage{amssymb}
\usepackage{mathrsfs}
\usepackage{natbib}
\usepackage{color}
\definecolor{upforestgreen}{rgb}{0.0, 0.55, 0.13}
\bibpunct{(}{)}{;}{a}{}{,}
\graphicspath{{./fig/}{./png/}}
\newcommand{\uuu}{{\bm u}}

\newcommand{\SSt}{\bm{\mathsf{S}}}
\newcommand{\SStij}{\mathsf{S}_{ij}}

\newcommand{\Equ}[1]{Equation~(\ref{#1})}
\newcommand{\Eqs}[2]{Equations~(\ref{#1}) and (\ref{#2})} 

\newcommand{\EQ}{\begin{equation}}
\newcommand{\EN}{\end{equation}}
\newcommand{\EQA}{\begin{eqnarray}}
\newcommand{\ENA}{\end{eqnarray}}

\newcommand{\pd}{\partial}

\newcommand{\mean}[1]{\overline{#1}}

\newcommand{\cP}{c_{\rm P}}
\newcommand{\cV}{c_{\rm V}}

\newcommand{\urms}{u_{\rm rms}}

\newcommand{\chiSGSz}{\chi_{\rm SGS}^{(0)}}
\newcommand{\chiSGSo}{\chi_{\rm SGS}^{(1)}}

\newcommand{\Hp}{H_{\rm p}}

\newcommand{\PraSGSz}{{\rm Pr}_{\rm SGS}^{(0)}}
\newcommand{\PraSGSo}{{\rm Pr}_{\rm SGS}^{(1)}}

\newcommand{\Rey}{{\rm Re}}

{}

\newcommand{\calR}{{\cal R}}
\newcommand{\FFFrads}{{\bm F}_{\rm rad}}
\newcommand{\FFFSGSs}{{\bm F}_{\rm SGS}}
\newcommand{\Fenth}{\mean{F}_{\rm enth}}

\newcommand{\nabad}{\nabla_{\rm ad}}

\newcommand{\tbot}{{\rm bot}}
\newcommand{\ttop}{{\rm top}}
\def\onethird{{\textstyle{1\over3}}}
\def\onehalf{{\textstyle{1\over2}}}

\newcommand{\Figp}[2]{Figure~\ref{#1}({#2})} 
\newcommand{\Figsp}[3]{Figures~\ref{#1}({#2}) and ({#3})} 
\newcommand{\Figu}[1]{Figure~\ref{#1}}
\newcommand{\Table}[1]{Table~\ref{#1}}
\begin{document}

\title{Extended subadiabatic layer in simulations of overshooting convection}

\author{Petri J.\ K\"apyl\"a$^{1,2,3}$, Matthias Rheinhardt$^{2}$, Axel Brandenburg$^{4,5,6,7}$, Rainer Arlt$^{1}$, Maarit J.\ K\"apyl\"a$^{3,2}$, \\Andreas Lagg$^{3}$, Nigul Olspert$^{2}$, and J\"orn Warnecke$^{3}$}
\affil{$^1$Leibniz-Institut f\"ur Astrophysik, An der Sternwarte 16,
  D-14482 Potsdam, Germany\\
  $^2$ReSoLVE Centre of Excellence, Department of Computer Science,
  P.O. Box 15400, FI-00076 Aalto, Finland \\
  $^3$ Max-Planck-Institut f\"ur Sonnensystemforschung,
  Justus-von-Liebig-Weg 3, D-37077 G\"ottingen, Germany\\
  $^4$NORDITA, KTH Royal Institute of Technology and Stockholm University,
  Roslagstullsbacken 23, SE-10691 Stockholm, Sweden\\
  $^5$Department of Astronomy, AlbaNova University Center,
  Stockholm University, SE-10691 Stockholm, Sweden\\
  $^6$JILA and Department of Astrophysical and Planetary Sciences,
  Box 440, University of Colorado, Boulder, CO 80303, USA\\
  $^7$Laboratory for Atmospheric and Space Physics,
  3665 Discovery Drive, Boulder, CO 80303, USA}
\submitted{Astrophys. J. 845, L23 (2017)}
\date{Received 2017 March 20; revised 2017 July 27; accepted 2017 July 29; published 2017 August 22}

  \begin{abstract}
    We present numerical simulations of hydrodynamic overshooting 
    convection in local Cartesian domains.
    We find that a substantial fraction of the lower part of the convection zone (CZ) is
    stably stratified according to the Schwarzschild criterion while
    the enthalpy flux is outward directed.
    This occurs when the heat conduction profile at the bottom
    of the CZ is smoothly varying, based either on a
    Kramers-like opacity prescription as a function of
    temperature and density or a static profile of a similar shape.
    We show that the subadiabatic layer arises due to nonlocal energy
    transport by buoyantly driven downflows in the upper parts of the
    CZ. Analysis of the force balance of the upflows and downflows
    confirms that convection is driven by cooling at the surface.
    We find that the commonly used
    prescription for the convective enthalpy flux being
    proportional to the negative entropy gradient does not hold in the
    stably stratified layers where the flux is positive.
    We demonstrate the existence of a
    non-gradient contribution to the enthalpy flux, which is estimated to be
    important throughout the convective layer.
    A quantitative analysis of downflows indicates a transition
      from a tree-like structure where smaller downdrafts merge into
      larger ones in the upper parts to a structure in the deeper
      parts where a height-independent number of strong downdrafts
      persist. This change of flow topology occurs when a
      substantial subadiabatic layer is present in the lower part of
      the CZ.
  \end{abstract}
  


\section{Introduction}

Convection plays a vital role in stellar activity by generating
turbulence that, together with the star's overall rotation,
leads to differential rotation \citep[e.g.][]{R89} and dynamo action
\citep[e.g.][]{KR80}.
Energy transport due to convection is important for almost all
stars during some stages of their evolution
\citep[e.g.][]{2012sse..book.....K}. Hence, a
proper parameterization
of convection
 is
crucial for stellar structure and evolution in
one-dimensional models.

Mixing length theory (MLT) continues to be a popular description
of stellar convection.
The formulation of MLT, as it is used today, goes back to the
seminal work of \cite{Vi53},
where the properties of convection are related to the local
value
of the superadiabatic gradient $\nabla-\nabla_{\rm ad}$,
with $\nabla=d\ln\overline{T}/d\ln\overline{p}$ and
$\nabla_{\rm ad}$ being the actual and adiabatic
logarithmic temperature gradients, respectively.
Here, 
$T$ and $p$ are temperature and pressure, respectively,
while overbars denote horizontal averaging.
Convection is supposed to occur only if the horizontally averaged
temperature stratification is superadiabatic, $\nabla>\nabla_{\rm ad}$,
which is equivalent to the Schwarzschild criterion
for convection, 
$d\,\overline{s}/dz<0$, where $s$ is the specific entropy.

The MLT has deeply influenced the way in which
three-dimensional \emph{ab initio} convection models are constructed:
often a fixed profile of radiative heat conductivity $K$ is
chosen, producing a superadiabatic
layer of fixed depth \citep[e.g.][]{HTM86}. Alternatively, the
static
thermodynamic
background state is taken from an 
MLT-based
stellar evolution model
\citep[e.g.][]{BMT11,KKMW16}.
In such setups, convection is driven at the largest scale
available. This is a possible cause
for
the discrepancy in the convective velocities at large
horizontal scales between simulations and time--distance helioseismology
\citep{HDS12}.
Smaller length scales could instead be imprinted 
by convection driven solely by the surface layers \citep{CR16},
which
leads to
a topology change of the downdrafts from a tree-like structure
 near the surface to
strong plumes penetrating 
into
deeper layers as cool \emph{entropy rain}; see \cite{Sp97}.
The latter
can
take part in the
convective flux in 
these
layers through a non-gradient
contribution known as 
\emph{Deardorff flux} \citep{De66,Br16}.

We present simulations
in which
we use either a physically motivated
heat conduction formulation based on a Kramers-like opacity \citep{BNS00} or
two
types of fixed heat conductivity profiles to study their effect on
the
structure of the convection zone (CZ).
Furthermore, we demonstrate the existence of a non-gradient contribution to the enthalpy
flux and use a quantitative analysis
to study the topology change of the downflow and upflow structures
in the simulations.

\section{The Model} \label{sect:model}

\subsection{Basic equations} \label{sect:basic}

We solve the equations of compressible hydrodynamics:
\begin{eqnarray}
\frac{D \ln \rho}{D t} &=& -\bm\nabla \cdot \uuu, \label{equ:dens}
\end{eqnarray}
\begin{eqnarray}
\frac{D\uuu}{D t} &=& {\bm g} -\frac{1}{\rho}(\bm\nabla p - \bm\nabla \cdot 2 \nu \rho \bm{\mathsf{S}}),\label{equ:mom} \\
T \frac{D s}{D t} &=& -\frac{1}{\rho} \left[\bm\nabla \cdot \left(\FFFrads + \FFFSGSs\right) \right] + 2 \nu \bm{\mathsf{S}}^2,
\label{equ:ent}
\end{eqnarray}
where $D/Dt=\pd/\pd t+\uuu\cdot\bm\nabla$ is the advective
derivative, $\rho$ is the density, $\uuu$ is the velocity,
$\bm{g}=-g\hat{\bm{e}}_z$, 
is the gravitational acceleration with $g>0$,
and $\nu$ is the constant kinematic viscosity.
$\FFFrads$ and $\FFFSGSs$ are the radiative and
subgrid scale (SGS) fluxes, respectively, and $\SSt$ is the traceless
rate-of-strain tensor with $\SStij=\onehalf
(u_{i,j}+u_{j,i})-\onethird\delta_{ij}\bm\nabla\cdot\uuu$. We
consider an optically thick, fully ionized gas. Thus, radiation is
taken into account through the diffusion approximation, and the ideal
gas equation of state $p=\calR\rho T$ applies,
where $\calR=\cP-\cV$ is the gas constant and $c_{\rm P,V}$ are the specific
heats at constant pressure and volume, respectively. 
The radiative
flux is given by $\FFFrads=-K\bm\nabla T$, where $K$ is the radiative
heat conductivity. It has either a fixed
profile $K(z)$ or it is a function of density and temperature,
$K(\rho,T)$, given by
$K=16\sigma_{\rm SB}T^3/3\kappa\rho$,
where $\sigma_{\rm SB}$ is the Stefan--Boltzmann constant and
$\kappa=\kappa_0(\rho/\rho_0)^a (T/T_0)^b$ is the opacity
with coefficient $\kappa_0$, exponents $a$ and $b$, and 
reference values of density and temperature, $\rho_0$, $T_0$.
These relations combine into
\begin{eqnarray}
K(\rho,T) = K_0 \, (\rho/\rho_0)^{-(a+1)} (T/T_0)^{3-b}.
\label{equ:Krad2}
\end{eqnarray}
Here we use $a=1$ and $b=-7/2$, corresponding to Kramers opacity law
for free-free transitions \citep{WHTR04}.

The radiative diffusivity 
$\chi=K/\rho\cP$ 
can vary by several
orders of magnitude as a function of depth in the Kramers opacity
case.
In order to keep the simulations numerically stable, we apply
additional turbulent SGS diffusivities in the entropy
equation,
\begin{eqnarray}
\FFFSGSs = -\rho T \left(\chiSGSz \bm\nabla \mean{s} + \chiSGSo \bm\nabla s'\right),
\label{equ:FSGS}
\end{eqnarray}
where $s'=s-\mean{s}$ is the fluctuation of specific entropy, and
$\chiSGSz$ acts on the mean entropy, is non-zero
only near the surface, while $\chiSGSo$ is constant
and acts on the entropy fluctuations.
We use the {\sc Pencil Code}\footnote{\url{https://github.com/pencil-code}}.

\subsection{Geometry, initial and boundary conditions}

The computational domain is rectangular with $-2 \leq(x,y)/d
\leq2$ and $-0.5\leq z/d\leq1$, where $d$ is the depth of the
initially isentropic layer.
The initial
stratification consists of two polytropic layers with 
indices $n_1=3.25$ in $-0.5\leq z/d<0 $ and $n_2=1.5$ in $0\leq z/d
\leq1$. The former is the same as in the special case where the
temperature gradient in the corresponding hydrostatic state is
constant; see \cite{BB14}.

The horizontal boundaries are periodic whereas the vertical boundaries
are impenetrable and stress free for the flow.
We set the
temperature gradient at the bottom 
according to 
${\pd_z T}=-{F_{\rm tot}}/{K_\tbot}$, 
where $F_{\rm tot}$ is a
fixed input flux and $K_\tbot$ is the value of the heat conductivity
at the bottom of the domain.
On the upper boundary we assume, for simplicity, a fixed
gradient of specific entropy such that
$({d}/{\cP})\,{\pd_z s}\equiv\widetilde{\pd_z s}=-10$.
This condition allows the density and temperature to vary locally.

\subsection{Control parameters and diagnostics}

Our models are fully defined by choosing the values of
$\nu$, $g$,
$a$, $b$, $\widetilde{\pd_z s}$, $F_{\rm tot}$, $K_0$, $\rho_0$, $T_0$, the SGS
Prandtl numbers
$\PraSGSz={\nu}/{\chiSGSz}(z/d=1)$ and $\PraSGSo={\nu}/{\chiSGSo}$, 
the $z$-dependent profile of $\chiSGSz$, and the initial normalized
pressure scale height at the surface, $\xi_0\equiv\Hp^{\rm
  (top)}/d={{\calR}\mean{T}_\ttop}/{gd}$.
The value of
$K_0$ is fixed by assuming $F_{\rm rad}=F_{\rm tot}$
at the bottom of the domain.
The normalized input 
flux is given by $F_n=F_{\rm tot}/\rho_\tbot
c_{\rm s,bot}^3$, where $\rho_\tbot$ and $c_{\rm s,bot}$ are
density and sound speed, respectively, at $z/d=-0.5$ in the
initial non-convecting state.
We also quote the Reynolds number,
$\Rey={\urms}/{\nu k_1}$,
where $\urms$
is the volume-averaged rms velocity and $k_1=2\pi/d$.

Dominant contributions to the mean vertical energy flux are
\begin{eqnarray}
  &\mean{F}_{\rm rad}  = -\mean{K}\,{\pd_z \mean{T}}, \label{eq:Frad}
  &\quad\mean{F}_{\rm kin}  = \onehalf \mean{\rho}\, \mean{\bm{u}^2 u_z}, \label{eq:Fkin} \\
  &\mean{F}_{\rm enth} = \cP \mean{(\rho u_z)' T'}, \label{eq:Fenth}
  &\quad\mean{F}_{\rm SGS}^{(0)} = - \chiSGSz \, \mean{\rho} \mean{T}\,{\pd_z \mean{s}}. \label{eq:FSGS}
\end{eqnarray}
The viscous energy flux is negligible.
No mean flows are generated, hence primes on $\bm{u}$ are dropped.

\section{Results} \label{sect:results}

Here, we describe the results of three simulations where the heat
conductivity is either based on Kramers' law (Run~K), or it has a
fixed profile that either coincides with the Kramers conductivity (P)
in the initial state of Run~K or a piecewise constant profile
(S); see, e.g., \cite{1994ApJ...421..245H}.

\begin{deluxetable}{cccccc}
\tabletypesize{\scriptsize} \tablecaption{Definition of the Zones.}
\tablewidth{0pt}
\tablehead{
\colhead{$\Fenth$} & \colhead{$\nabla-\nabad$} & \colhead{Zone} & \colhead{Label} &
\colhead{Lower Limit} & \colhead{Thickness}}
\startdata
$>0$       & $>0$ & \mbox{Buoyancy}  & \mbox{BZ} & $z_{\rm BZ}$ & $d_{\rm BZ}$ \\ 
$>0$       & $<0$ & \mbox{Deardorff} & \mbox{DZ} & $z_{\rm DZ}$ & $d_{\rm DZ}$ \\
$<0$       & $<0$ & \mbox{Overshoot} & \mbox{OZ} & $z_{\rm OZ}$ & $d_{\rm OZ}$ \\ 
$\approx0$ & $<0$ & \mbox{Radiative} & \mbox{RZ} &  $\cdots$    &  $\cdots$
\enddata
\label{tab:zonedef}
\end{deluxetable}
\vspace{-3mm}

\subsection{Revising the CZ structure}
\label{sec:structure}

\begin{figure}[t]
\centering
\includegraphics[width=0.49\textwidth]{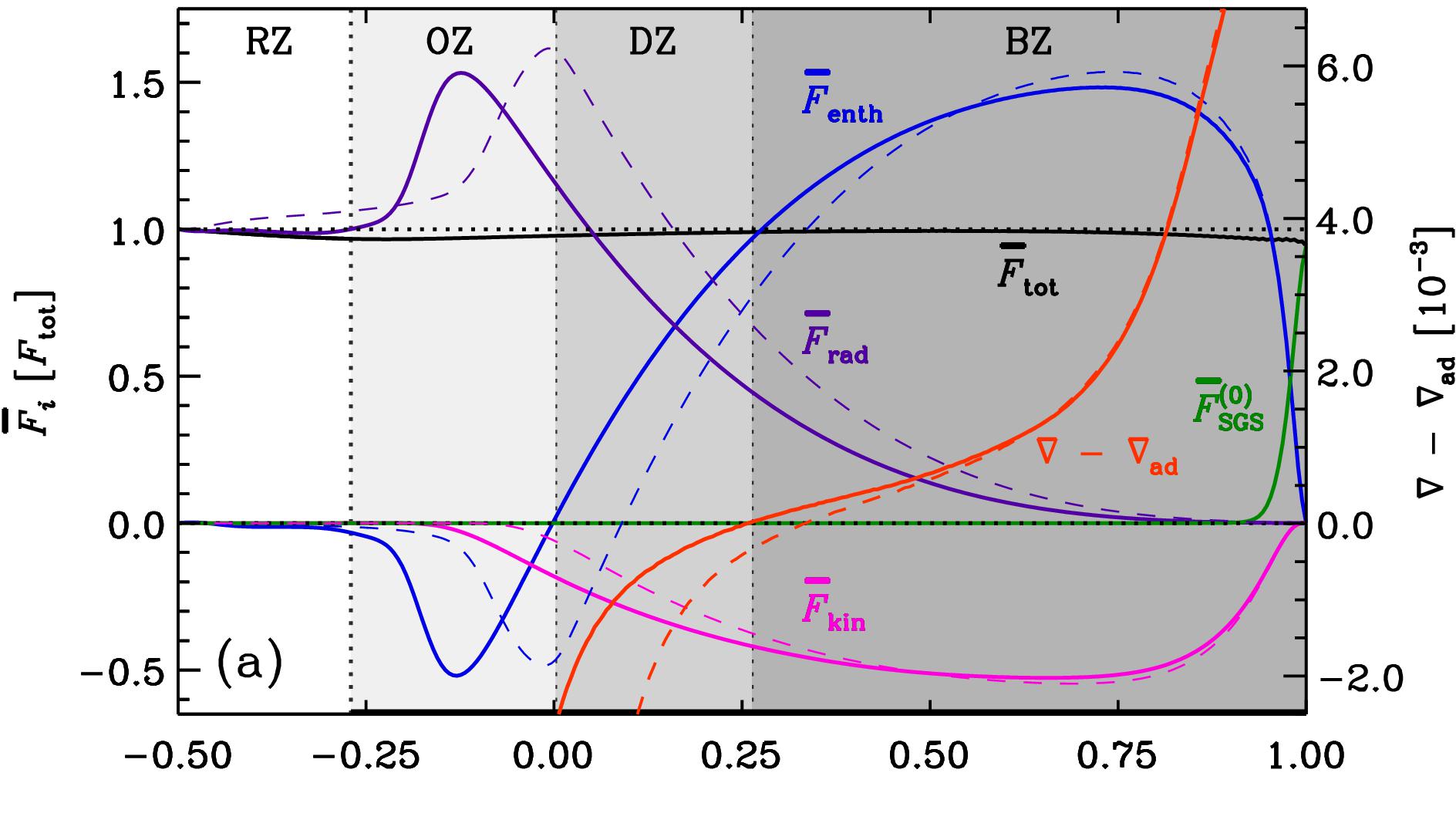}
\includegraphics[width=0.49\textwidth]{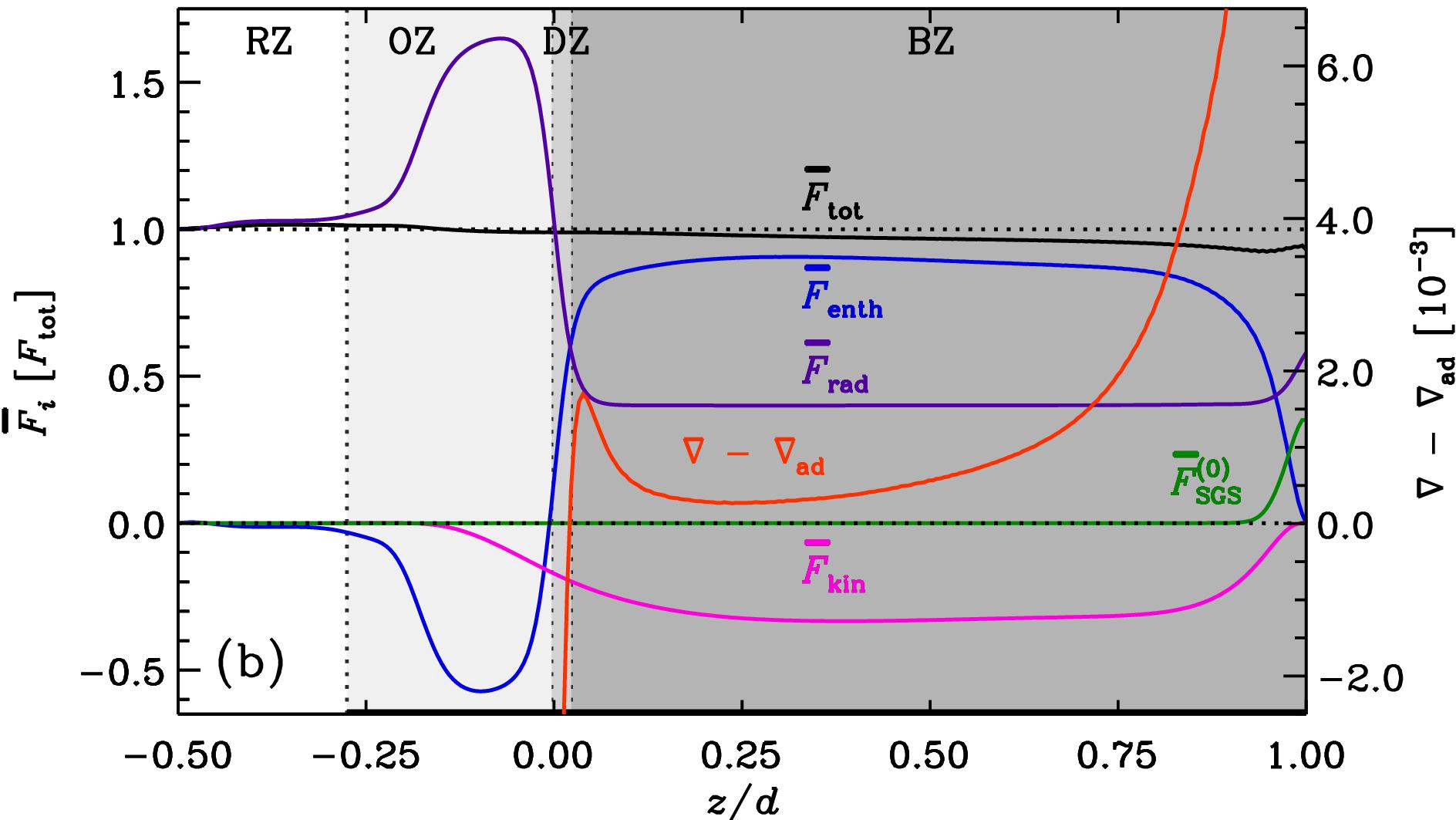}
\caption{
  Solid lines:
  radiative (purple), enthalpy (blue), kinetic energy (magenta), and
  SGS (green) fluxes for Runs~K 
    (a) and S (b).
  Red: 
  $\nabla-\nabla_{\rm ad}$. 
  Dashed lines in (a):
  corresponding data from Run~P.
  The thick horizontal lines 
  on the abscissae
  mark the extent
    of the MZ.
}
\label{fig:pfconv}
\end{figure}

As a basis for our analysis, we show in \Figsp{fig:pfconv}{a}{b}
the energy fluxes, defined by
\Eqs{eq:Frad}{eq:FSGS}, and the 
superadiabaticity, $\nabla-\nabla_{\rm ad}$.
Depending on the signs of enthalpy flux and
superadiabaticity, four different regimes and corresponding zones can
be identified; see Table~\ref{tab:zonedef}. The top three layers, 
BZ, DZ, and OZ,
are efficiently mixed while in the lowermost
(radiative) layer (RZ), mixing is inefficient.
In $z>z_{\rm BZ}$, we have $\nabla\geq\nabad$, whereas
in $z_{\rm DZ}<z<z_{\rm BZ}$ we have $\nabla\leq\nabad$
and yet $\mean{F}_{\rm enth}>0$.
In $z<z_{\rm DZ}$, we have $\mean{F}_{\rm enth}<0$, while in
$z<z_{\rm OZ}$ we also have $|\mean{F}_{\rm enth}|\le0.03F_{\rm tot}$.
The positions and thicknesses of the respective layers,
$d_{\rm BZ}$, $d_{\rm DZ}$, and $d_{\rm OZ}$
(see Table~\ref{tab:zonedef}), are listed in \Table{tab:runs}.
We refer to the union of BZ, DZ, and OZ
as the `mixed zone'
(MZ).
Our BZ and OZ are in the traditional parlance the
CZ and OZ, respectively, while the DZ has no
counterpart in the usual paradigm of convection.
Here, we consider the layers where $\Fenth>0$, i.e.\ the combination of BZ and 
DZ, as the revised CZ.

\begin{deluxetable}{ccccccccc}
\tabletypesize{\scriptsize} \tablecaption{Summary of the runs,
All with $288^3$ meshpoints.}
\tablecomments{
 Column `$K$':
 heat conduction scheme.
Remaining columns:
 depths and thicknesses of the zones;
  see Table~\ref{tab:zonedef}. 
  $\PraSGSz=0.5$, $\PraSGSo=1$, $F_n\approx2.1\cdot10^{-6}$, and 
  $\xi_0=0.054$.
}
\tablewidth{0pt}
\tablehead{
\colhead{Run} & \colhead{$K$}  & \colhead{$\Rey$} & \colhead{$z_{\rm BZ}$} & \colhead{$z_{\rm DZ}$} &
\colhead{$z_{\rm OZ}$} & \colhead{$d_{\rm BZ}$} & \colhead{$d_{\rm DZ}$} &
\colhead{$d_{\rm OZ}$}}
\startdata
K & \mbox{Kramers}     & $27$ & $0.26$ & $0.00$ & $-0.27$ & $0.74$ & $0.26$ & $0.27$ \\ 
P & \mbox{profile} & $25$ & $0.34$ & $0.10$ & $-0.19$ & $0.66$ & $0.24$ & $0.29$ \\ 
S  & \mbox{step}       & $26$ & $0.02$ & $0.00$ & $-0.28$ & $0.98$ & $0.02$ & $0.28$    
\enddata
\label{tab:runs}
\end{deluxetable}

We identify a DZ in Runs~K and P. In Run~K, $\mean{F}_{\rm enth}$
remains positive down to $z/d\approx 0$, although the superadiabaticity
already turns negative at $z/d=0.26$. Runs~P and K are similar, but
the BZ is somewhat shallower in P. This is a
consequence of the static profile of the heat conductivity as opposed
to the dynamic formulation of Run~K, where the depth of the MZ
is not known \emph{a priori}.
In Run~S with a fixed step profile for $K$,
the difference is more striking: even though the depths of the 
MZ
and CZ
are the same as in Run~K, the 
DZ
is negligibly
thin; see \Figu{fig:pfconv}(b).
This is due to the fact that the constant heat conductivity above 
$z/d=0$
forces radiative diffusion to transport a certain fraction of the flux
\citep{BCNS05} and the abrupt change of $K$
around $z=0$ prevents a smooth transition to a stable stratification 
beneath.

It is remarkable that
in Runs~K and P,
 the lower 
$\sim40\%$ of the 
MZ is stably stratified according to the
Schwarzschild criterion.
In these runs, the mixed, but stably stratified layer is roughly equally
divided 
into
DZ and OZ.
This is similar to the results of \cite{BNS00}, who were the first to
use a Kramers-based heat 
conductivity.
\cite{1993A&A...277...93R} used a temperature-dependent
heat conductivity in earlier two-dimensional simulations and
found a subadiabatic convective layer at the base of the CZ.
However, simulations of a $5M_\odot$ red giant star, employing a
heat conduction profile based on 
OPAL opacities,
did not indicate a DZ 
\citep{2013ApJ...769....1V}.
An extended subadiabatic convective layer was also reported by   
\cite{CG92} from a large-eddy simulation, although they applied a
prescribed
 step
function for the heat conductivity. However, their models had
low resolution and their subsequent works did not
mention
similar findings. More recent
studies reported subadiabatic convective layers in different contexts 
\citep[e.g.][]{2015ApJ...799..142T,2017arXiv170400817K,2017ApJ...843...52H}.

\begin{figure}[t]
\centering
\includegraphics[width=.5\textwidth]{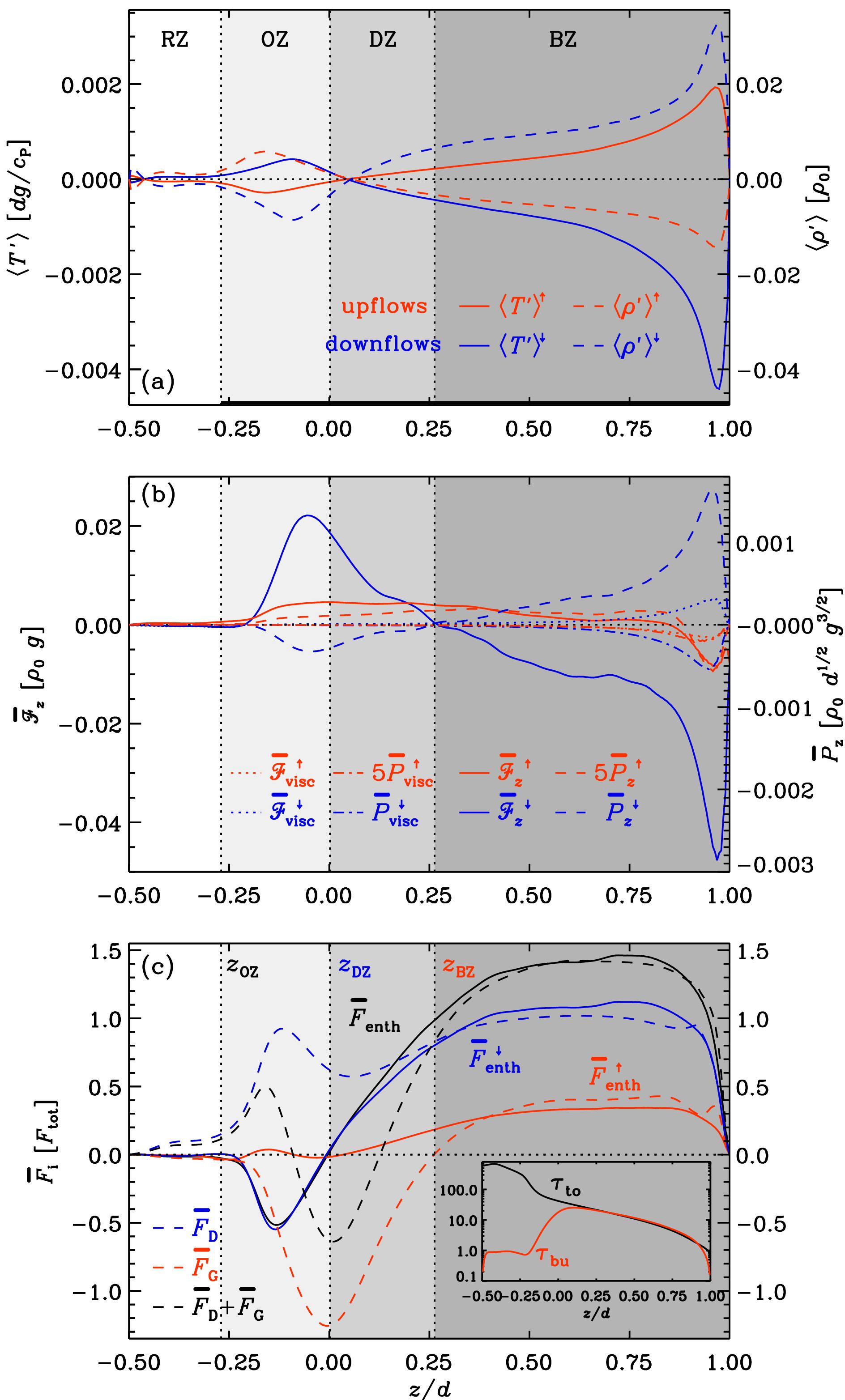}
\caption{(a) Temperature fluctuation (solid, left axis)
    and density fluctuation (dashed, right axis),
    averaged separately over upflows (red) and downflows (blue).
    (b) Horizontally
    averaged force $\mean{\mathscr{F}}_z=\mean{\rho
      Du_z/Dt}$ (solid lines, left axis), and the accelerating power
    $\mean{P}_z=\mean{u_z\mathscr{F}_z}$ of those forces (dashed,
    right axis). $\mean{P}_z^\uparrow$ is scaled up by a factor of
    five.
    $\mean{\mathscr{F}}_{\rm visc}$ and $\mean{P}_{\rm
        visc}$ are the corresponding viscous force and its power.
    (c) Averaged enthalpy flux (solid lines)
    along with parameterizations according to \Equ{equ:Fenth}
    (dashed). The inset shows $\tau_{\rm to}$ and $\tau_{\rm bu}$ as
    functions of depth in units of $\sqrt{d/g}$. 
    Data for Run~K.
\label{fig:pbuoy}
}
\end{figure}

\begin{figure*}[t]
\centering
\includegraphics[width=.49\textwidth]{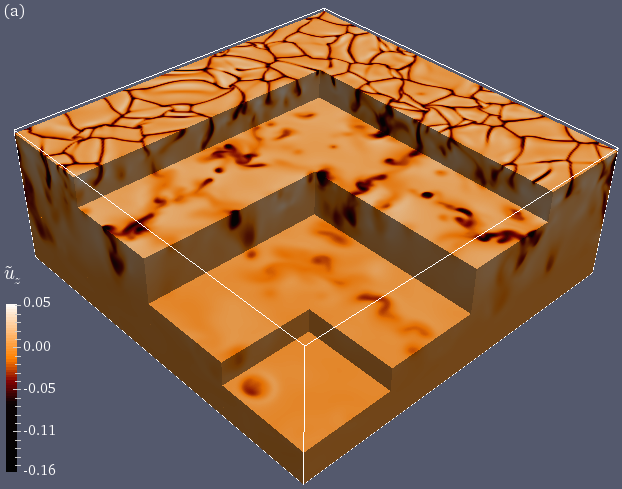}
\includegraphics[width=.49\textwidth]{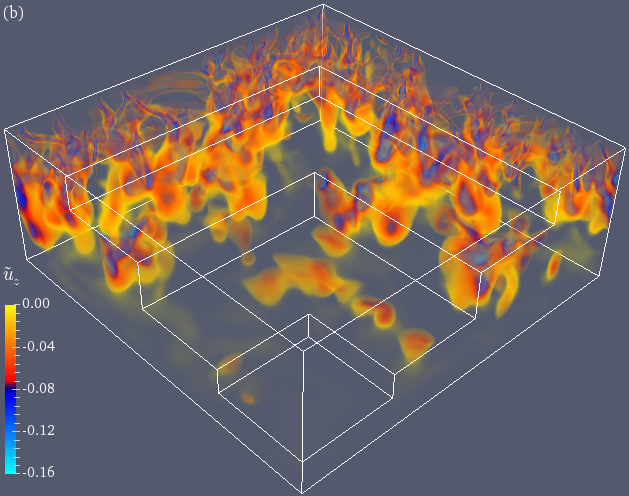}\vspace{.2cm}
\includegraphics[width=.33\textwidth]{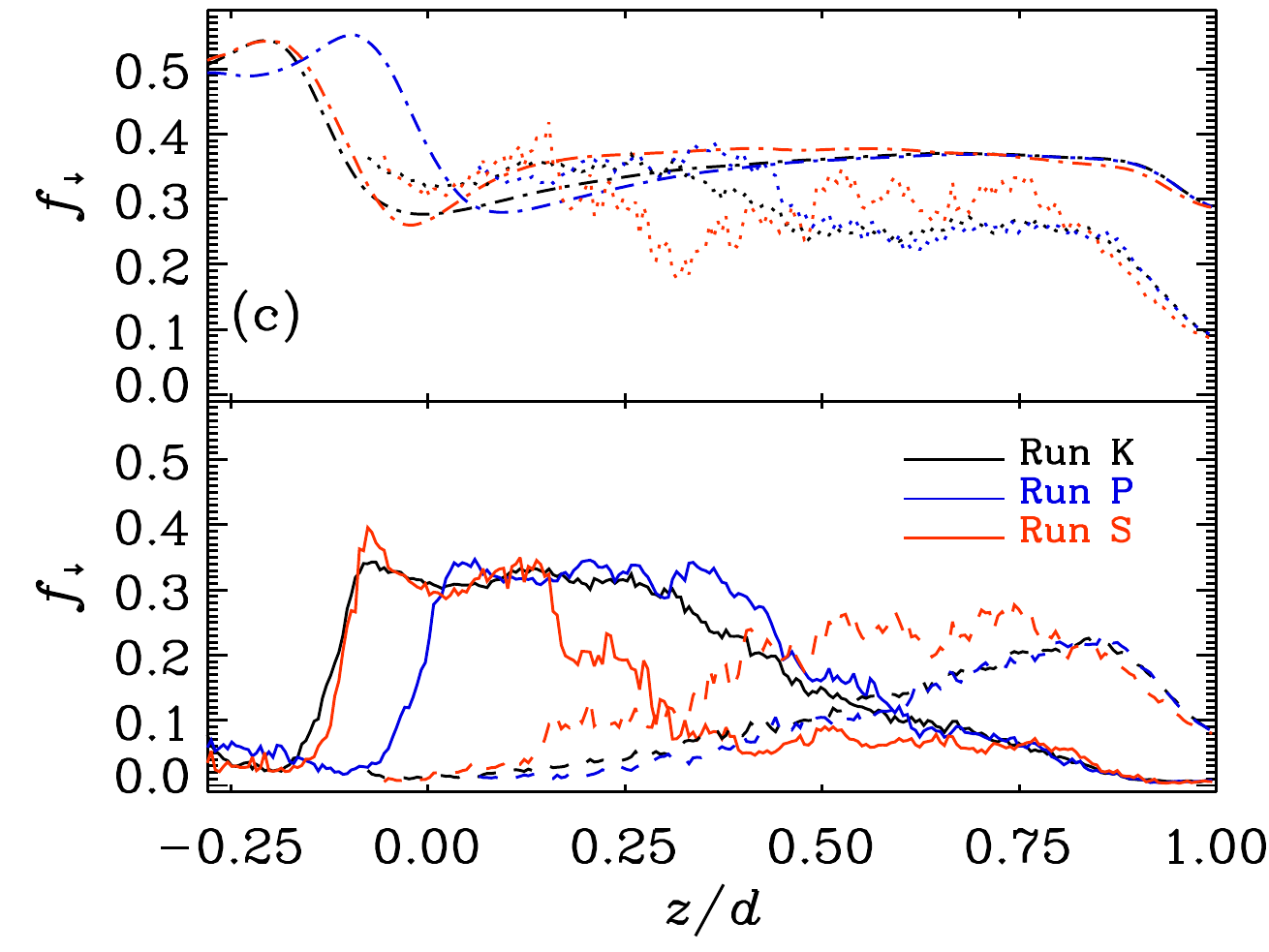}\includegraphics[width=.33\textwidth]{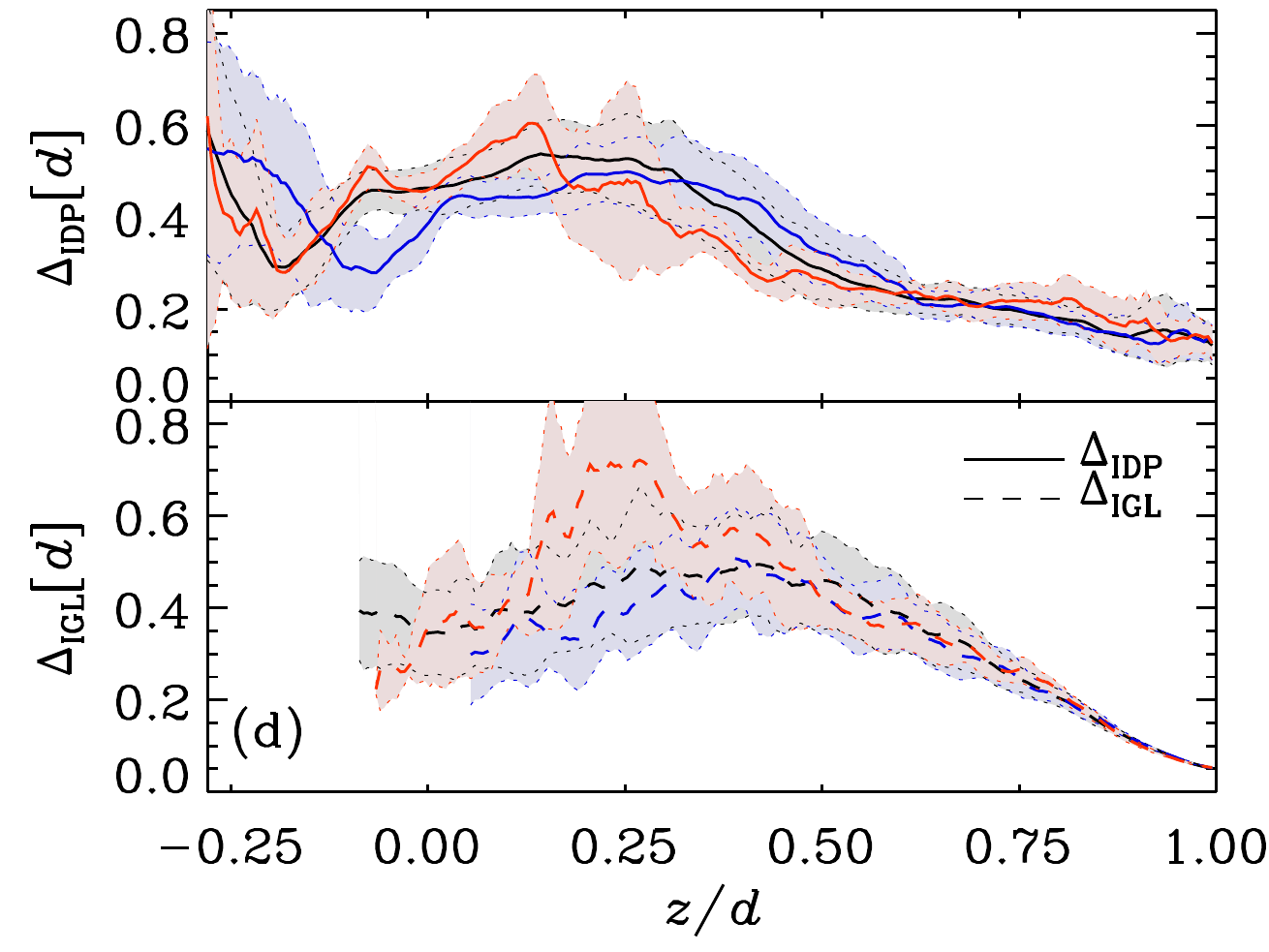}\includegraphics[width=.33\textwidth]{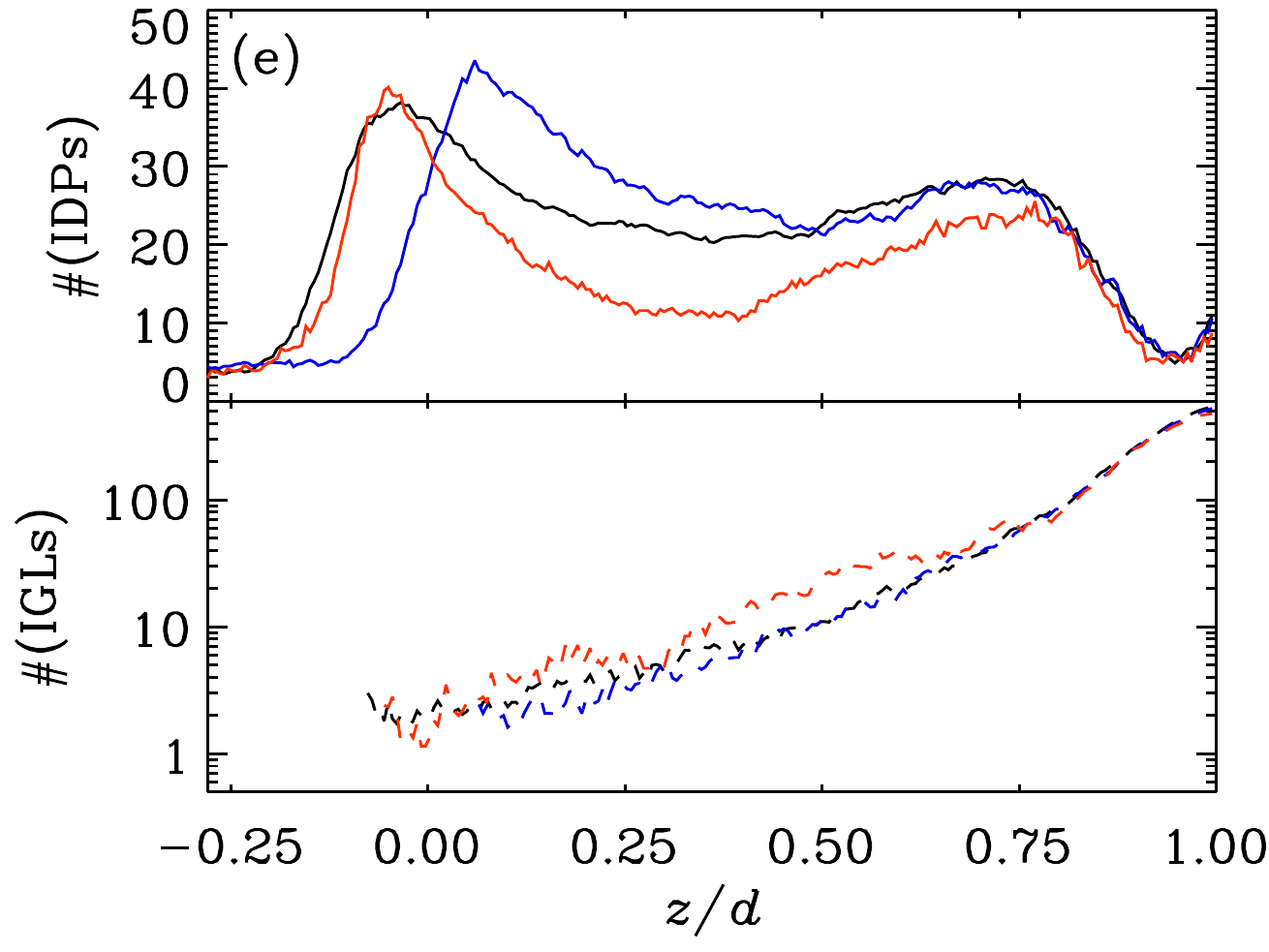}
\caption{(a) Vertical velocity and (b) volume rendering of downflows at the periphery and from depths
  $z/d=(0.98,0,63,0.13,-0.13)$ corresponding to a near-surface layer,
  and the middles of BZ, DZ, and OZ, respectively, in Run~K.
  Tildes refer to normalization
  with 
  $\tilde{u}_z=u_z/\sqrt{gd}$.
  (c) Filling
  factor of downflows (dash--dotted lines),
   a proxy filling factor for IDPs (solid),
   IGLs (dashed, only occurring at $z/d\gtrsim 0.0$)
  and their sum 
  (dotted), 
  (d) average widths of 
  IDPs and IGLs
  with the standard deviations from the temporal evolution
  (shaded areas), and (e) their numbers as functions of $z$
  from Runs~K (black), P (blue), and S (red).}
\label{fig:boxplot_sf288b2}
\end{figure*}

\subsection{Why does an extended DZ exist?}

We show in \Figu{fig:pbuoy}(a) for Run~K that the fluid in the upflows
(downflows) is lighter (heavier) and warmer (cooler) than average
in almost all of the DZ,
indicating that both contribute to positive $\Fenth$.
Furthermore, \Figu{fig:pbuoy}(b) shows that, in the BZ, the total force
$\mean{\mathscr{F}}_z=\mean{\rho Du_z/Dt}$ is negative for the
downflows and changes sign at $z_{\rm BZ}$, while for the upflows,
$\mean{\mathscr{F}}_z$ is positive everywhere except very near the surface.
The associated power, $\mean{P}_z=\mean{u_z\mathscr{F}_z}$ (blue dashed),
shows that the downflows gain energy mostly near the
surface while the upflows (red dashed) are accelerated throughout the MZ except
near the surface.
The viscous force is non-negligible only near the surface (dotted and
dash--dotted).
We speculate
that there the upflows are decelerated by viscous momentum exchange with the downflows.

Based on these data, we interpret the DZ as an
{\it overshooting phenomenon}, but with an upward enthalpy flux, and
with
the resulting stable stratification
being
nearly adiabatic.
We explain the appearance of the DZ such that
cool fluid elements that originate near the surface are 
accelerated downward by gravity and gain enough momentum to
penetrate the convectively stable layer beneath the BZ. There they are
progressively decelerated and heated up. If this
proceeds fast enough, fluid elements, having kept a sufficient part of
their momentum, can continue moving downward, but now with an {\it
excess of entropy} thus forming the OZ.
The upflows in the stably stratified OZ cannot be due to
convective instability, but are driven by the pressure excess 
exerted by
the matter 
in
the downflows.
In the Schwarzschild stable DZ, the upflows are lighter than their surroundings,
which is an important property of the DZ; see Figure~1(d) of \cite{Br16}.
However, the force on the downflows is {\it decelerating};
see \Figp{fig:pbuoy}{b}.
Therefore, the upflows in the DZ are not buoyancy-driven.
Instead, we argue that they are pressure driven, just like in the OZ.

We conclude that the Deardorff layer is associated with
\emph{nonlocal} transport of
heat 
as the 
downdrafts
of cool matter originating from the
strongly cooled surface
propagate
not only through the BZ, but further on to the bottom of the DZ.
This process is called \emph{entropy rain}
which characterizes 
stellar convection as being driven
by radiative cooling
at the surface \citep{SN89}.

In an attempt to quantify the different contributions to the enthalpy
flux, we compare
the numerical results for $\Fenth$ with a mean-field parameterization that
takes into account the non-gradient
contribution introduced by \cite{1961JAtS...18..540D,De66}.
Here, we use the expression derived by \cite{Br16}:
\begin{equation}
\Fenth^{\rm MF} = \tau_{\rm red}\, \mean{\rho}\, \mean{T}
\big(g\, \overline{{s'}^2}/\cP - \mean{u_z^2}\, \pd_z \mean{s}\big)
\equiv \mean{F}_{\rm D} + \mean{F}_{\rm G},
\label{equ:Fenth}
\end{equation}
where $\tau_{\rm red}$ is a reduced relaxation time, taking into
account radiative cooling and turbulent energy transport.  The first
term, $\mean{F}_{\rm D}$, describes the non-gradient Deardorff flux,
which is positive irrespective of the sign of the entropy gradient,
whereas the latter term, $\mean{F}_{\rm G}$, is the traditional mean-field description of
the (gradient) enthalpy flux. In \Figp{fig:pbuoy}{c}, we show $F_{\rm
  D}$, $F_{\rm G}$, and their sum for Run~K, assuming that $\tau_{\rm
  red}(z)=c_\tau \tau(z)$, where
$\tau(z)=\min(\tau_{\rm to},\tau_{\rm bu})$ is the minimum of the
convective turnover time $\tau_{\rm to}=\Hp/\urms$ and the buoyancy
timescale $\tau_{\rm bu}=(c_{\rm P}/g)(\mean{u_z^2}/\mean{{s'}^2})^{1/2}$,
and $c_\tau$ is a free parameter, for the best fit set to 0.73.
\Figp{fig:pbuoy}{c} shows that \Equ{equ:Fenth} provides a good description in
the BZ and 
even
suggests that the Deardorff term is the dominant contribution
to the heat transport. However, the expression in
\Equ{equ:Fenth} breaks down in the DZ and OZ.
Further,
we separate $\Fenth$ in \Figp{fig:pbuoy}{c} into contributions from upflows
($\Fenth^\uparrow$) and downflows ($\Fenth^\downarrow$) for
Run~K.
The downflows dominate the enthalpy flux with $\Fenth^\downarrow
\approx 3 \Fenth^\uparrow$.
We find that gradient and Deardorff contributions match
$\Fenth^\uparrow$ and $\Fenth^\downarrow$, respectively, in the BZ.
However, $F_{\rm D}$ and $F_{\rm G}$ contain contributions from both
upflows and downflows, and the correspondence to $\Fenth^\uparrow$ and
$\Fenth^\downarrow$ is likely coincidental. This conjecture is
supported by \Figp{fig:pbuoy}{b}, which suggests that the downflows feel
the local entropy gradient.
The generality of these results
will be investigated elsewhere using wider parameter studies.

Two recent studies \citep{2017arXiv170400817K,2017ApJ...843...52H}
have reported subadiabatic layers from convection simulations. The
former authors studied 
``weakly non-Boussinesq''
 convection in spherical coordinates
where the 
radial dependence
of the superadiabatic temperature gradient
of the background state was varied. They found that a
subadiabatic layer appeared in regions where convective transport was
efficient (or radiative diffusion inefficient). They also studied the
contributions of upflows and downflows separately and found that the
upflows in these cases contributed to \emph{downward} flux of heat.
This is qualitatively different to our cases where $\Fenth^\uparrow$
is always nearly zero (OZ, lower part of DZ) or positive (upper part
of DZ, BZ). In their case, the upflows are clearly pressure driven even
in the bulk of the CZ -- in contrast to our simulations.

The study of \cite{2017ApJ...843...52H} is a close parallel to ours in
that a 
density-stratified, 
initially piecewise polytropic setup was
used to study overshooting in fully compressible simulations, although
with magnetic fields. The main
difference to our runs is that our density contrast is larger
($37$ in the CZ of Run~K) compared to about 6--7 in
\cite{2017ApJ...843...52H}.
Furthermore, he used a fixed profile of $K$ that is smoother than in
our Run~S, but steeper than in our Run~P. This results in a similar
subadiabatic layer at
the base of the CZ as in our Runs~K and P. Moreover,
\cite{2017ApJ...843...52H} analyzed the vertical force balance (his
Figure~13) and came to the conclusion that the total buoyancy force switches
sign roughly at
the same location as the entropy gradient. 
We conclude that the mechanism
forming a subadiabatic layer in the runs of \cite{2017ApJ...843...52H} is
very likely the same as in our models.

\subsection{Structure of convection}

Given the appearance of an extended DZ, we analyze
 the flow
 in detail
to find out whether 
its
topology in the DZ is altered in comparison
to the BZ.
We adopt the approach of \cite{Br16}, where the structure of
convection is characterized by the number and filling factor of
downflows (cf.\ his Figure~2). We
have developed a dedicated algorithm (to be reported on elsewhere)
to detect isolated downflow plumes (IDPs) and intergranular lanes
(IGLs), compute their sizes and numbers,
and thereby their filling factors for such an analysis.
In Case~I of \cite{Br16},
corresponding to forest-like downflow structures,
number,
size, and filling factor of the downflows are 
independent
of depth. 
His
MLT description 
including the Deardorff flux
is closest to 
his
Case III, with a tree-like structure, where the filling
factor is constant, but the number (size) of the downflows decreases
(increases).

In \Figsp{fig:boxplot_sf288b2}{a}{b}, we show representative 
patterns of the vertical velocity of Run~K
from
BZ,
DZ, and
OZ.
They are
qualitatively similar to those found in numerous other studies of
stratified convection
\citep[e.g.][]{SN89,SNGBS09,HRY14,KBKKR16,KKMW16}:
upwelling granules with downflows along
a network of connected
IGLs
near the surface.
Deeper down, the cellular structure
disintegrates and IDPs appear. In the DZ and OZ
only a few IDPs survive in the midst of much larger scale
upflows.

The filling factors
$f_\downarrow$ of 
all
downflows, 
and of IGLs and IDPs separately
(for the latter two
a proxy 
using
average sizes and numbers to 
ease comparison with panels
(d), (e)),
are shown in
\Figu{fig:boxplot_sf288b2}(c).
We find that $f_\downarrow$
of all downflows
is almost independent of depth in the redefined CZ (= BZ + DZ) for all runs.
The filling factors of IDPs and IGLs reveal
that for Run S in the bulk of the CZ ($0.3\lesssim z/d \lesssim 0.8$), the dominant
IGLs have roughly a constant filling factor while their size increases and their number
decreases, consistent with the tree-like structure of Case III.
For Runs K and P, the structure of convection is clearly distinct from S, with the
IGL network starting to disappear at much smaller depths, and the IDPs already taking
over at $z/d \approx 0.6$.

After the IDPs take over as the dominating structure of convection,
we observe another difference between Run S and Runs K and P. In the latter cases,
after a smooth transition, the IDP filling factor attains a constant value,
while their size is constant and the number is mildly
  increasing. These data correspond most closely to Case I. This
holds roughly at depths $0\lesssim z/d \lesssim 0.35$,
encompassing the DZ and the lower parts of the BZ.
Thus, the tree-like picture
roughly holds until $z/d \approx 0.35$, below which a
depth-independent number of IDPs persist. In Run S, the IDP filling
factor also
tends to a constant in between $-0.1\lesssim z/d \lesssim 0.2$, but this
is accompanied by an increase of the number and decrease of the size of IDPs,
incompatible with Cases~I and III. 
These results suggest that the structure of the downflows in the
  DZ and the bottom part of the BZ is qualitatively different from
  that in the upper parts of the BZ (forest-like instead of
  tree-like).

\section{Conclusions}

We have shown that, when a smoothly varying heat conduction profile is used,   
a substantial part of the lower CZ is weakly
subadiabatic although $\Fenth>0$.
A smooth
transition can also be expected to occur in deep stellar interiors            
where
 a Kramers-based conductivity
 is 
valid.
Furthermore,
 with such a heat conduction
 law, the depth of the
CZ is an outcome of the simulation and cannot be
determined \emph{a priori}.

We have shown that the subadiabatic layer can still lead to
an upward enthalpy flux due to downflows
bringing low entropy material from near the surface to the stably
stratified layers below. We also found that the upflows in the
\emph{overshoot zone} are driven by the pressure excess due to the
matter brought down by the downflows.
Except for the lowermost parts, the ascending matter in
  the Deardorff layer is lighter than the surroundings.
  Yet, we argue that also in the Deardorff layer the upflows are
  pressure driven.
  There is no buoyant acceleration, and the downflows are instead decelerated in
  accordance with the Schwarzschild criterion.
  Our results confirm
that convection is highly nonlocal and 
driven by cooling at the
surface resulting in cool entropy rain.
The traditional mean-field expression of the enthalpy flux fails in
the subadiabatic part of the CZ and a non-gradient term
is required. Our work demonstrates the existence of such a contribution,
introduced by \cite{1961JAtS...18..540D} and applied to stellar MLT by \cite{Br16},
for the first time
from numerical simulations.
A 
geometric analysis shows a transition from a tree-like to
a forest-like structure in the deep parts of the CZ.
Energetic considerations reveal
that the downflows 
provide
the dominant contribution to
the enthalpy flux everywhere in the CZ. Furthermore, as they are
largely responsible for driving the upflows, their importance for the
overall convective structure is indeed crucial.

The current simulations may have too low resolution to
fully
capture the driving of strong
downflows near the surface 
so that
their effect in more realistic conditions can be even more
  pronounced. This can lead to a further reduction of the depth of the
  BZ, which, in conjunction with the topology change, could alleviate
the discrepancy between
helioseismic and 
simulation-based
estimates of convective velocities
\citep{MFRT12}. 
Furthermore, our
results are obviously at odds with MLT, which is widely used in
stellar structure models, calling for more advanced one-dimensional
models
 \citep[see,
  e.g.][]{1999ApJ...526L..45K,2015AN....336...32S}.
Another implication of a subadiabatic layer above the base of the CZ
comes from potentially breaking the Taylor--Proudman balance of
the solar rotation profile \citep{Re05}.
These questions will be addressed elsewhere.

\begin{acknowledgements}
  The authors thank an anonymous referee for his/her constructive
  comments and criticism on the manuscript.
  The simulations were performed using the supercomputers hosted by
  CSC -- IT Center for Science Ltd.\ in Espoo, Finland, who are
  administered by the Finnish Ministry of Education. Special Grand
  Challenge allocation NEOCON is acknowledged. Financial support from
  the Academy of Finland grant No.\ 272157 to the ReSoLVE Centre of
  Excellence (P.J.K., M.R., M.J.K., N.O.) is acknowledged.
J.W. acknowledges funding from the People Programme (Marie Curie
Actions) of the European Union's Seventh Framework Programme
(FP7/2007-2013) under REA grant agreement No.\ 623609.
\end{acknowledgements}

\bibliography{paper}


\end{document}